# *A versatile platform for gas-phase molecular polaritonics*

Adam D. Wright,[1] Jane C. Nelson,[1] and Marissa L. Weichman[1,a]

**AFFILIATIONS**

[1]*Department of Chemistry, Princeton University, Princeton, New Jersey, 08544, USA*

[a]Author to whom correspondence should be addressed: weichman@princeton.edu

**ABSTRACT**

Strong cavity coupling of gas-phase molecules will enable studies of benchmark chemical processes under strong light-matter interactions with a high level of experimental control and no solvent effects. We recently demonstrated the formation of gas-phase molecular polaritons by strongly coupling the bright $v_3$, $J = 3 \rightarrow 4$ rovibrational transitions of methane ($CH_4$) to a Fabry-Pérot optical cavity mode inside a cryogenic buffer gas cell. Here, we further explore the flexible capabilities of this infrastructure. We show that we can greatly increase the collective coupling strength of the molecular ensemble to the cavity by increasing the intracavity $CH_4$ number density. In doing so, we can tune from the single-mode coupling regime to the multimode coupling regime in which many nested polaritonic states arise as the Rabi splitting approaches the cavity mode spacing. We explore polariton formation for cavity geometries of varying length, finesse, and mirror radius of curvature. We also report a proof-of-principle demonstration of rovibrational gas-phase polariton formation at room temperature. This experimental flexibility affords a great degree of control over the properties of molecular polaritons and opens up a wider range of simple molecular processes to future interrogation under strong cavity-coupling. We anticipate that ongoing work in gas-phase polaritonics will facilitate convergence between experimental results and theoretical models of cavity-altered chemistry and physics.

**I. INTRODUCTION**

Polaritons are mixed light-matter states which arise from the resonant exchange of energy between an optically bright transition in matter and the confined photonic mode of an optical cavity.[1-5] The spectroscopic signature of such strong light-matter coupling is the formation of distinct peak splittings in the cavity transmission spectrum, separated in frequency by $\Omega_R$, the vacuum Rabi splitting.[6, 7] Originally identified in cold atoms[8] and in semiconductors,[9] polaritons have recently become a topic of great excitement in chemistry, given the observations of altered reaction rates in cavity-coupled solution-phase molecules.[1-3, 5, 10-17] The prospect of achieving selective control over chemical bonds using vibrational strong coupling (VSC) is particularly enticing.[2, 11] However, theoretical understanding of the mechanisms underlying vibrational polariton chemistry remains lacking, hindered in part by the complexity of the reactive systems involved.[11, 12, 18-21] Examining polariton behavior in the gas phase may provide the experimental means to study clean, isolated systems for which theoretical treatment and hence understanding may be more tractable.[3, 14, 22-25]

We recently reported the first demonstration of vibrational strong coupling in gas-phase molecules by coupling a rovibrational transition of methane ($CH_4$) cooled within a cryogenic buffer gas cell to a Fabry-Pérot optical cavity.[26] It is experimentally non-trivial to realize strong light-matter coupling of vibrational transitions in gas-phase molecules: low gas-phase number densities and oscillator strengths act to reduce the accessible collective Rabi splitting, which scales with the square root of the number density of cavity-coupled molecules and linearly with the transition dipole moment.[6, 7] However, accessing the strong coupling regime requires only that $\Omega_R$ exceed both the molecular and cavity mode linewidths. We were able to reach this regime in methane by working with a cold, dense gas sample, which served to minimize the rovibrational partition function and narrow the Gaussian Doppler broadening of the molecular lineshape, such that $\Omega_R$ could be resolved with modest molecular number density.[26]

Here, we explore the extended capabilities of this gas-phase molecular polariton platform, showcasing the flexibility with which we can reconfigure the gas sample and cavity parameters. We continue to work with methane as a stable spherical top molecule which features isotropic cavity-coupling strength regardless of its orientation in the lab frame. We target cavity-coupling of individual rovibrational transitions in methane's $\nu_3$ asymmetric C−H stretching band, which is optically bright and has been previously used for schemes in vibrational mode-selective chemistry.[27, 28] At sufficiently high $CH_4$ number densities we observe multiple nested polariton features whose appearance we rationalize in terms of each molecular transition coupling to the manifold of cavity modes as $\Omega_R$ approaches the cavity mode spacing. We additionally report VSC in a room-temperature gas for the first time, providing a more widely accessible route to engineer strongly cavity-coupled gases. Finally, we explore a range of cavity parameters, including length, finesse, and mirror radius of curvature, and discuss the possibilities that the resulting control over polariton linewidths and per-molecule coupling strengths may enable for future studies of cavity-altered chemistry and physics. This versatile setup will provide a new testbed for studying molecular polaritonics with quantum-state-specific cavity coupling and no complications from solvent.

## II. METHODS

In our experiments, we introduce a gaseous sample into a temperature-controlled cell surrounded by a feedback-stabilized Fabry-Pérot optical cavity enclosed within a vacuum chamber. We use continuous-wave mid-infrared spectroscopy to record cavity transmission spectra, which we simulate using classical optics. These methods are described in detail in our previous work.[26] Here, we provide a brief overview with an emphasis on the experimental modifications specific to this study.

### A. Experimental methods
#### 1. Cryogenic buffer gas cell

Our homebuilt cryogenic buffer gas cell, depicted in Fig. 1, draws inspiration from instruments used for cavity-enhanced molecular spectroscopy.[29, 30] The chamber contains a cell mounted to the base of a liquid nitrogen (LN$_2$) dewar; gas flowing into the cell thermalizes to the cell wall temperature via collisions.[26, 30, 31] When the chamber is evacuated and the dewar is filled with LN$_2$, the system reaches a base pressure of $1.1\times10^{-5}$ torr.

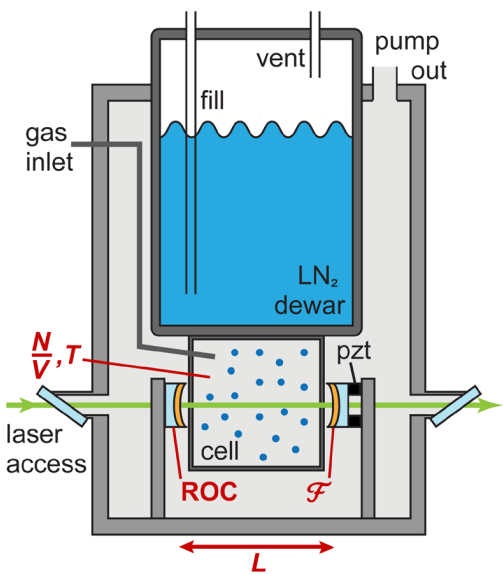

**FIG. 1.** Cryogenic buffer gas cell apparatus used for cavity-coupling a gas-phase molecular sample. The gas cell is housed inside a vacuum chamber and enclosed within a Fabry-Pérot optical cavity to achieve *in situ* strong coupling. The apparatus permits control over the parameters indicated in red: the number density ($N/V$) and temperature ($T$) of the intracavity gas, the length ($L$) and finesse ($\mathcal{F}$) of the cavity, and the radius of curvature ($ROC$) of the plano-concave cavity mirrors. Adapted with permission from A. D. Wright, J. C. Nelson and M. L. Weichman, J. Am. Chem. Soc. **145**, 5982 (2023). Copyright 2023 American Chemical Society.

Here, we make use of two schemes to introduce the molecular gas of interest to the cell. In our first scheme, the same as that used in our prior work,[26] we use pre-cooled helium (Airgas) as an inert buffer gas to thermalize CH$_4$ (Airgas) to the cell wall temperature. The helium gas is cooled by passage through a line mounted to the outer surface of the LN$_2$ dewar before flowing through an annular slit inlet which surrounds the front aperture of the cell. Room temperature CH$_4$ gas is introduced to the cell through this front aperture. In this configuration, we use mass flow controllers (Alicat Scientific) to control the flow of both helium and CH$_4$ gases, fixing the helium flow to 100 sccm and varying the CH$_4$ flow between 0 and 15 sccm. Under these conditions, the translational temperature of methane reaches 120 K, based on a measured Doppler linewidth of 180 MHz full-width at half-maximum (fwhm).

In this work, we introduce a second gas flow scheme in order to target higher intracavity $CH_4$ number densities. In this new scheme, depicted in Fig. 1, we route the $CH_4$ directly through the high flow limit (1000 sccm) mass flow controller, pre-cooled gas line, and inlet used for helium in the original configuration. In doing so, we use a single gas line and no helium buffer gas for this scheme. Buffer gas cooling with an inert gas like helium is generally the most efficient means for universal molecular cooling, but the volatility and low condensation point of methane render the use of a secondary buffer gas not strictly necessary at $LN_2$ temperatures. Under these high $CH_4$ flow conditions, we report a translational temperature of $115 \pm 3$ K for $CH_4$ based on a measured Doppler linewidth of $176 \pm 2$ MHz fwhm, with uncertainties arising from variation depending on the flow rate and dewar fill level. We attribute this lower temperature to more efficient cooling afforded by prolonged direct contact of $CH_4$ with the dewar in contrast to incomplete thermalization through collisions with He buffer gas. We obtain room temperature data at high gas flow rates with the same gas line configuration, but with an empty $LN_2$ dewar. For a given methane flow rate, the intracavity pressure depends on both the gas temperature and the gas line configuration. We therefore recover the intracavity methane number density ($[CH_4]$) for each experiment by fitting a simulation to the measured cavity transmission spectra (see Sec. II B).

### *2. Fabry-Pérot optical cavities*

To achieve strong coupling, we construct in-vacuum, two-mirror Fabry-Pérot optical cavities around our cryogenic cell. We mount the cavity mirrors on either side of the gas cell in standard optical mirror mounts (Thorlabs Polaris) clamped to the floor of the chamber. We fabricate our cavity mirrors by depositing gold layers on the concave faces of plano-concave 1" diameter $CaF_2$ substrates (EKSMA Optics) using either an electron beam evaporator (Angstrom Engineering Nexdep) or metal sputterer (Angstrom Engineering EvoVac). We use gold coatings with nominal thicknesses of 5, 11, and 20 nm to achieve mirror reflectivities ($R$) of ~74%, ~88%, and ~93%, respectively. We use these mirrors to construct optical cavities with finesses ($\mathcal{F}$) of ~10, ~25, and ~43, given by $\mathcal{F} = \pi\sqrt{R}/(1-R)$.[32] Mirror reflectivities determined from fits to measured cavity transmission spectra (see Sec. II B) are consistent with reflectivity measurements obtained directly with an infrared microscope (Thermo Scientific Nicolet iN10 MX).

To achieve near-confocal cavity geometries, we use −8.36 cm or −6.27 cm radius of curvature (ROC) mirror substrates, and space the mirrors by a cavity length matched to their ROC. For the non-confocal cavity described in Sec. III E, we use ROC −21.69 cm substrates and a cavity length $L \sim 8.36$ cm. We have fabricated two cryogenic gas cells of length 7.62 cm and 5.72 cm to fit snugly within both longer and shorter cavity lengths, ensuring the cavity mode volumes are filled with thermalized molecules.

We actively stabilize the cavity length to ensure consistent cavity coupling conditions by performing a side-of-line lock to a cavity fringe at a wavelength near 1550 nm, using a stable metrology-grade continuous wave laser (RIO ORION) as the frequency reference. The error signal from this lock is fed back onto a piezo electric ring chip (Thorlabs) glued to one of the cavity mirrors in order to actuate the cavity length. More details on this cavity stabilization scheme are provided in our earlier report.[26]

### 3. Spectroscopy

We use a continuous-wave distributed feedback interband cascade laser (ICL, Nanoplus) to perform mid-infrared spectroscopy of $CH_4$ gas and optical cavity transmission. The ICL features a sub-10 MHz instantaneous linewidth, and a central wavelength that can be swept over 3262 to 3278 nm (3065 to 3050 cm$^{-1}$). This laser is designed specifically to target our desired $v_3$ $J = 3 \rightarrow 4$ transitions of methane which lie near 3.27 μm. In order to record cavity transmission spectra, we mode-match the ICL beam to the cavity using plano-convex $CaF_2$ lenses. We measure the intensity of transmitted light on an HgCdTe detector (Kolmar) and record a spectrum by sweeping the ICL wavelength. We measure $CH_4$ transmission spectra by removing the cavity mirrors from the chamber. We calibrate the ICL frequency using a room temperature methane vapor cell as an absolute frequency reference and a precision germanium etalon (Light Machinery) as a relative frequency reference. The ICL power incident on the input window of the chamber during our measurements is relatively low at 430 μW. For all the systems studied herein, our tests revealed no influence of probe laser power on the structure of the cavity transmission spectra, as shown in the representative power dependent measurements given in Fig. S1 of the supplementary material (SM). More details on our spectroscopy and frequency calibration schemes are provided in our earlier report.[26]

### B. Simulation of optical cavity transmission spectra

We simulate optical cavity transmission spectra using the classical expression for the fractional intensity of light transmitted through a Fabry-Pérot cavity with two identical mirrors:[26, 32, 33]

$$\frac{I_T(v)}{I_0} = \frac{T^2 e^{-\alpha(v)L}}{1+R^2 e^{-2\alpha(v)L}-2Re^{-\alpha(v)L}\cos\left(\frac{4\pi L n(v)v}{c}\right)} + \text{p.s.} \quad (1)$$

Here $L$ is the cavity length, $R$ and $T$ are the reflectivity and transmission coefficients for a single cavity mirror, $\alpha(v)$ and $n(v)$ are the frequency-dependent absorption coefficient and refractive index of the intracavity medium, $c$ is the speed of light, and $v$ is the laser frequency. When we simulate near-confocal cavities, we include a "phase shifted" term (designated p.s.) in Eq. 1 which accounts for the twice-as-dense mode spacing of the near-confocal cavity as compared to a non-confocal cavity.[32] This phase shifted term is identical to the first term apart from the addition of a phase shift of π within the argument of the cosine function.

We calculate $\alpha(v)$ and $n(v)$ for $CH_4$ under our experimental conditions beginning with $\sigma(v)$ absorption cross section data from HITRAN.[34] We use PGOPHER software[35] to process $\sigma(v)$ and correctly account for internal molecular temperature, Gaussian Doppler broadening, and Lorentzian pressure broadening. We then obtain the absorption coefficient $\alpha(v)$ via $\alpha(v) =$ $\sigma(v) \times [CH_4]$, and in turn calculate the imaginary component of the refractive index $\kappa(v)$ via $\kappa(v) = \alpha(v) c/(4\pi v)$. Finally, we obtain the real component of the refractive index, $n(v)$, from $\kappa(v)$ using the Kramers-Kronig relation implemented in MATLAB.[36]

We manually fit Eq. (1) to the experimental cavity transmission spectra with $L$, $[CH_4]$, $R$ and $T$ as fitting parameters. As shown in Fig. S2 in the SM, we find that the simulations are improved by scaling Eq. (1) by a factor of $e^{-\alpha(v)d}$ to capture the absorption of light by extracavity methane along a beampath of length $d$ inside the chamber on either side of the cavity mirrors. This spectral filtering effect is particularly evident when the $CH_4$ absorption lines are broad, as is the case at high $CH_4$ flow rates and at room temperature.

## III. RESULTS AND DISCUSSION

We now detail the behavior of the polaritonic cavity transmission under tuning of several different cavity and gas parameters. We access a variety of regimes that are important to characterize before pursuing future studies of intracavity chemistry and polariton dynamics in gas-phase molecules. As in our previous work,[26] we focus on cavity-coupling individual rovibrational transitions in the fundamental $v_3$ asymmetric C−H stretching band of methane, and in particular the $J = 3 \rightarrow 4$ multiplet lying close to 3.27 μm (3058 cm$^{-1}$).[37] We plot the experimental ~115 K transmission spectrum of $CH_4$ in this region in the upper trace of Fig. 2a with the three components of the $J = 3 \rightarrow 4$ transition designated by their symmetry labels. We principally target resonance between a single longitudinal cavity mode and the $A_2(0)$ transition of $CH_4$ at 3057.687423 cm$^{-1}$, whose frequency we use throughout this work as a reference point corresponding to 0 MHz. We also cavity-couple to the nearby $F_2(0)$ and $F_1(0)$ transitions at 3057.726496 cm$^{-1}$ (1171 MHz) and 3057.760735 cm$^{-1}$ (2198 MHz), respectively.

### A. Accessing larger Rabi splittings and multimode cavity coupling at high molecular number densities

In our initial study of gas-phase rovibrational strong coupling,[26] we observed Rabi splittings on the order of 450 MHz (0.015 cm$^{-1}$) for the $v_3$, $J = 3 \rightarrow 4$ $A_2(0)$ transition of $CH_4$, at 120 K and with $[CH_4] \sim 3.5 \times 10^{15}$ cm$^{-3}$. This collective cavity-coupling strength is orders of magnitude smaller than the relevant molecular vibrational and rotational energy scales as well as the Rabi splittings achievable in solution-phase systems. We therefore set out to investigate what coupling conditions could be accessed with higher intracavity $CH_4$ number density. With the experimental modifications described in Sec. II A, we can now access $CH_4$ flow rates up to 1000 sccm while maintaining ensemble temperatures near 115 K. In these experiments, we work with a near-confocal cavity with finesse $\mathscr{F} \sim 25$ and $L \sim 8.36$ cm.

These low-temperature, high-flow cavity transmission data are shown in Fig. 2. The colored traces in Fig. 2a depict the experimental cavity transmission spectrum measured for intracavity $CH_4$ number densities spanning $[CH_4] = 0$ to $1.6 \times 10^{17}$ cm$^{-3}$, corresponding to flow rates of 0 to 1000 sccm. In each trace, the cavity length is locked such that one cavity fringe remains on resonance with the target $A_2(0)$ transition. Fig. 2b shows simulated cavity transmission spectra as a function of [CH$_4$] under similar resonance conditions. Experimentally, when [CH$_4$] is increased from 0 to $2.6\times10^{15}$ cm$^{-3}$, the resonant cavity mode splits into a pair of peaks. When this peak

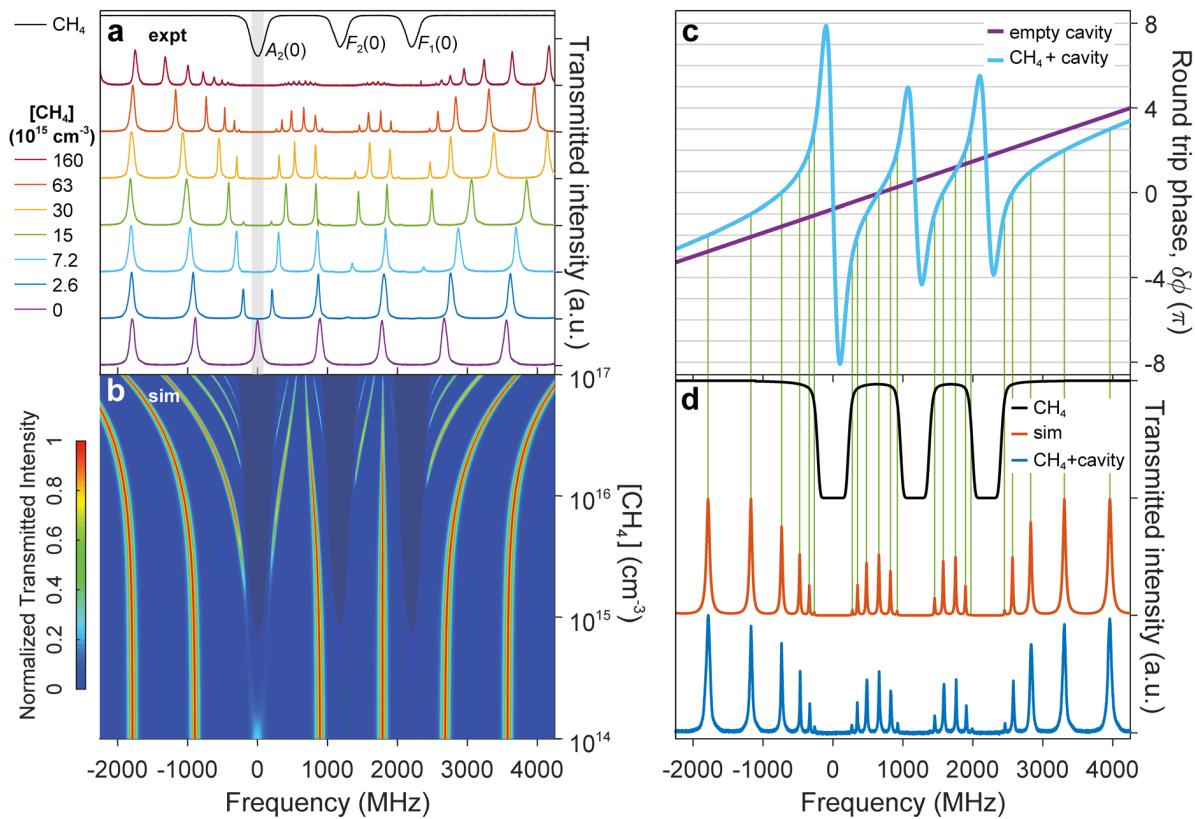

**FIG. 2. (a)** Experimental transmission spectra of a near-confocal Fabry-Pérot cavity with finesse $\mathscr{F} \sim 25$ and length $L \sim 8.36$ cm, under various intracavity number densities [CH$_4$] at ~115 K (colored traces). Successive spectra are vertically offset for clarity, as is the experimental transmission spectrum of CH$_4$ acquired at ~115 K and [CH$_4$] = $2.0 \times 10^{15}$ cm$^{-3}$ (black). In all traces, the cavity mode highlighted in gray is kept locked on resonance with the CH$_4$ $v_3$ $J = 3\rightarrow4$ $A_2(0)$ transition at 3057.687423 cm$^{-1}$. The frequency axis (300 MHz $\approx$ 0.01 cm$^{-1}$) is referenced with this transition corresponding to 0 MHz. **(b)** Simulated cavity transmission spectra as a function of [CH$_4$] demonstrating the expected appearance of nested multimode polaritons at high CH$_4$ flow rates. **(c)** Simulated round-trip phase ($\delta\phi = \frac{4\pi L n(v) v}{c}$) for an electromagnetic wave propagating in the cavity when it is empty (purple) and when it contains [CH$_4$] = $6.3 \times 10^{16}$ cm$^{-3}$ (cyan), which corresponds to a flow rate of ~300 sccm under our conditions. Horizontal gray lines indicate when $\delta\phi$ reaches an integer multiple of $\pi$, at which point a resonance will appear in the cavity transmission spectrum (vertical green lines). **(d)** Cavity transmission spectra obtained experimentally (blue) and simulated (orange) with [CH$_4$] = $6.3 \times 10^{16}$ cm$^{-3}$ (300 sccm gas flow). The simulated single-pass transmission spectrum ($e^{-\alpha(v)L}$) of CH$_4$ for these conditions is additionally plotted in black. The polaritonic cavity resonances which fall within the strongly absorbing regions are extinguished and do not appear in the cavity transmission spectra.

splitting exceeds the 75 MHz fwhm cavity linewidth and 176 MHz fwhm absorption linewidth of the $A_2(0)$ transition, the system enters the strong coupling regime. We therefore ascribe these split cavity transmission features to the formation of polaritons for CH$_4$ flow rates above ~1×10$^{15}$ cm$^{-3}$. As we increase the CH$_4$ flow rate, the splitting between the two polariton peaks grows with the expected collective $\Omega_R \propto [CH_4]^{1/2}$ scaling.[3, 26]

Intriguingly, for [CH$_4$] = 1.5×10$^{16}$ cm$^{-3}$ and above, we see the appearance of first one pair, then several pairs of weaker features nested within the original polariton peaks. Higher-frequency cavity modes, spaced ~895 MHz apart, also begin to couple off-resonantly to the nearby $F_2(0)$ and $F_1(0)$ transitions at higher [CH$_4$], resulting in an increasingly congested cavity spectrum. We observe a maximum notional Rabi splitting of 3540 MHz (0.118 cm$^{-1}$) for the $A_2(0)$ transition at [CH$_4$] = 1.6×10$^{17}$ cm$^{-3}$. For such large [CH$_4$], the extraction of the Rabi splitting is complicated by spectral congestion of the upper polariton bands near the $F_2(0)$ and $F_1(0)$ transitions. We therefore estimate the Rabi splitting under these conditions by tracking the position of the lowest-frequency polariton and doubling its frequency shift from the original molecular transition. It also becomes difficult to determine the cavity resonance conditions under high flow rates. Our experimental heuristic to identify the "on-resonance" condition is to lock the cavity length such that the weakest visible pair of nested polaritonic features appear symmetrically about the $A_2(0)$ transition.

These nested polariton features do not result from the bunching up of existing cavity fringes, but rather arise from either side of each strong CH$_4$ transition as the gas flow is increased. The appearance of these new features may be understood by considering the argument of the cosine function in Eq. (1), $\delta\phi = \frac{4\pi L n(\nu)\nu}{c}$, where $\delta\phi$ represents the round-trip phase accumulated by electromagnetic waves travelling in the cavity. Cavity transmission maxima occur for the frequencies at which the cosine term reaches its maximum value of +1, which serves to minimize the denominator in Eq. (1). In general, optical resonances occur whenever $\delta\phi = 2\pi m$, for integer $m$. Here, with the inclusion of the phase-shifted term for a confocal Fabry-Pérot cavity, optical resonances are supported for $\delta\phi = \pi m$. For a constant $n(\nu) = n_0$ in the absence of strong intracavity absorption, this resonance condition gives rise to the expected regular confocal cavity mode spacing of $\frac{c}{4Ln_0}$.[32] In the presence of a strong intracavity absorber, however, $n(\nu)$ exhibits a strongly dispersive lineshape near each molecular transition which can allow the $\delta\phi = m\pi$ phase condition to be met multiple times for a given value of $m$.

We illustrate these resonance conditions for our system in Fig. 2c, where we plot $\delta\phi$ for the cavity both when it is empty (purple trace) and under strong coupling with intracavity [CH$_4$] = $6.3 \times 10^{16}$ cm$^{-3}$ (cyan trace). Fig. 2d illustrates the corresponding strongly-coupled experimental and simulated cavity transmission spectra. As indicated by the vertical green lines in Fig. 2cd, the transmission spectra indeed exhibit maxima at the frequencies where $\delta\phi$ reaches an integer multiple of $\pi$ (grey horizontal lines in Fig. 2c). Transmission peaks which fall too close to a given CH$_4$ transition are suppressed by the strong molecular absorption.

Similar cavity transmission spectra featuring nested polaritonic features have been previously reported under VSC in molecular liquids[38, 39] and under electronic strong coupling in an atomic gas.[40] These additional features appear when the Rabi splitting approaches the original cavity mode separation, which has been termed the "superstrong coupling" regime.[40] These features can also be understood in terms of multimode cavity-coupling[41, 42] under conditions where the molecular absorption is strong enough to couple off-resonantly to neighboring longitudinal cavity modes.[38, 40] Within this picture, one may consider that the resulting mixed light-matter states have contributions from more than one photonic mode.[39] By accessing this multimode coupling regime, our apparatus may thus be well-suited to test theories of polariton and dark mode delocalization and their potential impacts on cavity chemistry.[43] On the other hand, we can easily increase the spacing between cavity modes by shortening the cavity (see Sec. III B) and so avoid the multimode coupling regime if desired.

More broadly, the wide tunability of intracavity molecular number density in our system will facilitate future studies of polariton chemistry spanning the weak to strong coupling regimes as a function of $\Omega_R$. In our current apparatus, we can vary [$CH_4$] over four orders of magnitude from $1.5 \times 10^{13}$ cm$^{-3}$ to $1.6 \times 10^{17}$ cm$^{-3}$ by changing the gas flow rate, constrained only by the mass flow controllers we have on hand. We work with flowing gas here in order to permit gas thermalization to cryogenic temperatures, as a static gas volume would ultimately equilibrate with the room temperature vacuum chamber walls. With the extension of gas-phase molecular polaritonics to room temperature samples, considerably higher static gas pressures with correspondingly higher Rabi splittings may become accessible, as discussed further below in Sec. III C.

### B. Simultaneous cavity-coupling of multiple molecular transitions

We next set out to investigate the prospects for simultaneous coupling of multiple molecular transitions to neighboring longitudinal cavity modes, as has been demonstrated in solution-phase molecules under VSC.[38, 44] We decrease the cavity length to $L \sim 6.27$ cm, such that the cavity mode spacing approaches the 1171 MHz frequency separation between the $A_2(0)$ and $F_2(0)$ transitions of the $v_3$, $J = 3 \rightarrow 4$ manifold. As in Sec. III A, we maintain a near-confocal cavity geometry with finesse $\mathscr{F} \sim 25$, and cool $CH_4$ to ~115 K without the use of any He buffer gas. In Fig. 3a, we show the dependence of the experimental cavity transmission spectrum on intracavity [$CH_4$] with the cavity length locked such that two adjacent cavity fringes couple to the $A_2(0)$ and $F_2(0)$ transitions. As we increase [$CH_4$], both fringes split into pairs of polariton peaks. At the highest number densities, multi-peaked spectra emerge characteristic of multimode coupling, as described above in Sec. III A. Fig. 3b shows the experimental cavity transmission dispersion plot for [$CH_4$] = $7.2 \times 10^{15}$ cm$^{-3}$ (10 sccm gas flow). We step the cavity length such that each fringe scans the full cavity mode separation. As expected, this dispersion plot exhibits the avoided crossings characteristic of strong coupling as consecutive cavity fringes pass through resonance with each $v_3$, $J = 3 \rightarrow 4$ $CH_4$ transition, and demonstrates that neighboring molecular transitions can be coupled simultaneously for cavity detunings between 0 and −400 MHz. The observed Rabi splitting is largest for the $A_2(0)$ transition, which has the largest absorption cross section. As we show in Fig. S3 of the SM, the Rabi splittings achieved in this shorter cavity are identical to those observed in the $L \sim 8.36$ cm system.

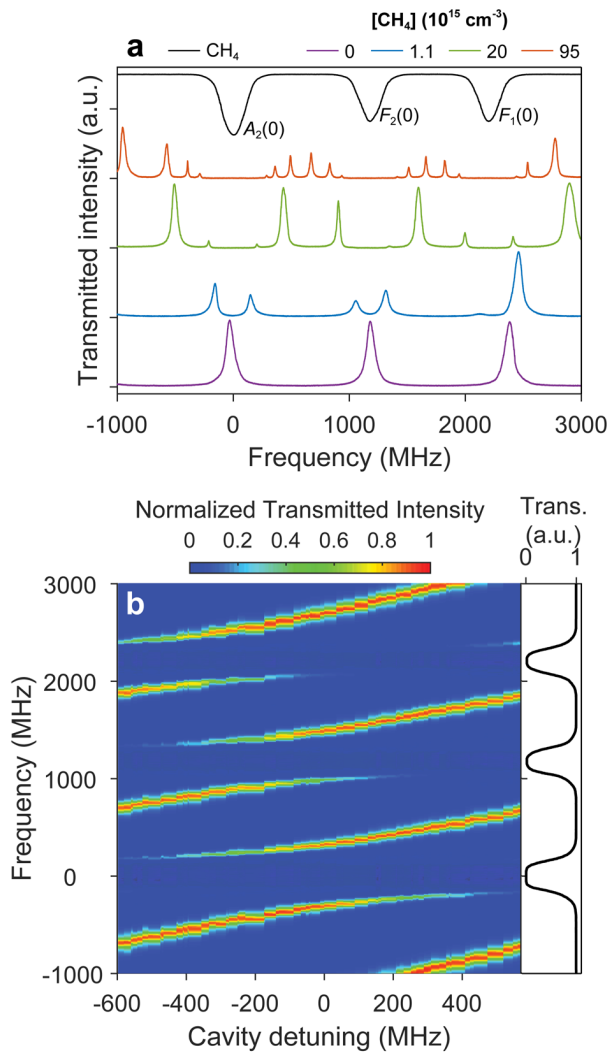

**FIG. 3. (a)** Experimental transmission spectra of a near-confocal Fabry-Pérot cavity of finesse $\mathcal{F} \sim 25$ and length $L \sim 6.27$ cm, for various intracavity number densities [CH$_4$] with a molecular temperature of ~115 K. Successive spectra are vertically offset for clarity, as is the experimental transmission spectrum of CH$_4$ acquired at ~115 K and [CH$_4$] = $2.0 \times 10^{15}$ cm$^{-3}$ (black). In all traces, the cavity length was locked such that adjacent cavity modes were near-resonant with the CH$_4$ $\nu_3$ $J = 3 \rightarrow 4$ $A_2(0)$ and $F_2(0)$ transitions. The frequency axis is referenced with the $A_2(0)$ transition at 3057.687423 cm$^{-1}$ corresponding to 0 MHz. **(b)** Dispersion plot of the experimental cavity transmission spectra for [CH$_4$] = $7.2 \times 10^{15}$ cm$^{-3}$ (10 sccm gas flow). The cavity mode frequencies are systematically detuned by varying the cavity length. The experimental transmission spectrum of CH$_4$ acquired at ~115 K and [CH$_4$] = $7.2 \times 10^{15}$ cm$^{-3}$ is plotted for reference in black.

The control our apparatus offers over cavity length yields several possibilities for future studies. The length of our open cavity (and hence its mode spacing) can be varied continuously over orders of magnitude, without the challenges imposed by the compressible spacers used in solution-phase microcavities. Precisely matching the cavity mode structure to manifolds of molecular transitions could facilitate the engineering of unusual light-matter states. As we can already see for the highest flow rate in Fig. 3a, the multimode polaritonic peaks formed between the methane lines feature contributions from both molecular transitions as well as multiple photonic modes. In this case, the neighboring $CH_4$ transitions arise out of lower rotational states that share the same $J = 3$ angular momentum quantum number, differing only in the projection of $J$ on the molecular frame. In future work, it should be feasible to achieve simultaneous multi-state coupling in a system where we can create polaritons with, for instance, mixed angular momentum character. It is conceivable that cavity-mediated rovibrational state mixing could enable new energy transfer pathways, just as coupling multiple material transitions to a single cavity mode has been observed to impact energy transfer in condensed-phase systems.[45, 46]

### C. Extending strong coupling to room temperature gas-phase molecules

Having explored strong coupling in cryogenically cooled molecules, we now extend our investigation to room temperature systems which may offer a more widely accessible route to the study of gas-phase polaritons. Retaining the near-confocal cavity of length $L \sim 6.27$ cm and finesse $\mathscr{F} \sim 25$ from Sec. III B, we lock the cavity length such that a cavity mode is on resonance with the target $v_3$, $J = 3 \rightarrow 4$ $A_2(0)$ transition of $CH_4$. We then flow room temperature (295 K) $CH_4$ gas into the cell. As the $CH_4$ intracavity number density is increased we again see the emergence of split peaks in the cavity transmission spectrum (Fig. 4a). The room temperature sample features a higher threshold for entering the strong coupling regime than the cryogenic sample due to the increased broadening of the molecular lineshape. The Rabi splitting ultimately exceeds both the resonant cavity mode linewidth (87 MHz fwhm) and the room-temperature Doppler-broadened $CH_4$ linewidth (282 MHz fwhm), confirming that the system resides in the strong coupling regime. Lorentzian pressure broadening is small compared to the Doppler broadening under our conditions, reaching 8 MHz fwhm at the highest flow rates. We plot simulations of these cavity transmission spectra in Fig 4b which achieve excellent agreement with the experimental data and confirm that our system is capable of generating gas-phase vibrational polaritons at room temperature.

Comparison of the room temperature cavity transmission data in Fig. 4a with the cryogenic results in Fig. 2b reveals that $\Omega_R$ is substantially smaller at room temperature than at 115 K for the same [$CH_4$]. To give a concrete example, we observe $\Omega_R = 888$ MHz (0.03 cm$^{-1}$) at 295 K and $\Omega_R = 1675$ MHz (0.06 cm$^{-1}$) at 115 K for [$CH_4$] = $4.2 \times 10^{16}$ cm$^{-3}$. The dependence of the achievable Rabi splitting on temperature derives largely from the changing rovibrational partition function, which increases from $Q = 143$ at 115 K to $Q = 587$ at 295 K.[34] Accordingly, the absorption coefficients for our target transitions decrease as an increasingly small fraction of the molecular population resides in the correct low-lying rovibrational state at higher temperatures. For example, at 115 K, 11 % of the molecular ensemble is found in the lower level of the $v_3$, $J = 3 \rightarrow 4$ $A_2(0)$ transition, compared to 4 % at 295 K. The comparative experimental ease of room temperature gas-phase strong coupling therefore comes at the cost of the vast majority of molecules being found in the wrong initial rovibrational state to couple to the cavity. Given this high background of uncoupled molecules, it may be easier to resolve cavity-altered behavior in cryogenic samples.

Regardless of these considerations, working with static room temperature intracavity samples will allow future studies of cavity-coupling with considerably higher intracavity gas pressures while avoiding the experimental complications associated with efficiently cooling such samples. For methane at room temperature and atmospheric pressure ($[CH_4] \sim 2.5 \times 10^{19}$ $cm^{-3}$), Lorentzian pressure broadening will broaden the molecular lineshape to 0.16 $cm^{-1}$ fwhm. Under these conditions, the three $v_3$, $J = 3 \rightarrow 4$ $CH_4$ components convolve into a single transition that simulations predict could be cavity-coupled to achieve a Rabi splitting of ~1.1 $cm^{-1}$. Herrera and

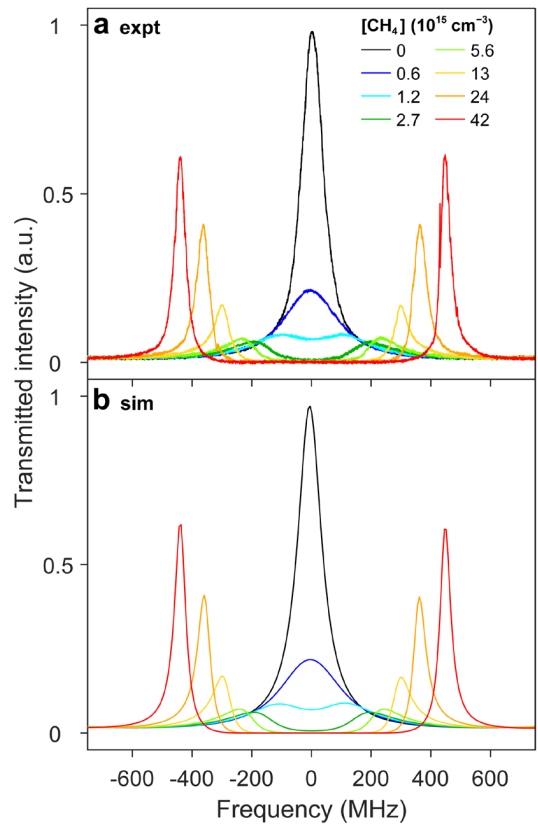

**FIG. 4. (a)** Experimental transmission spectra of a near-confocal Fabry-Pérot cavity of finesse $\mathscr{F} \sim 25$ and length L ~ 6.27 cm for various intracavity number densities [$CH_4$] flow rates at room temperature (295 K). In all traces, a cavity mode is kept locked on resonance with the $CH_4$ $v_3$ $J = 3 \rightarrow 4$ $A_2(0)$ transition at 3057.687423 $cm^{-1}$. The frequency axis (300 MHz $\approx 0.01$ $cm^{-1}$) is referenced with this transition corresponding to 0 MHz. The traces are normalized against the intensity of a cavity fringe far from resonance. **(b)** Simulated cavity transmission spectra obtained by fitting the corresponding experimental traces. The legend applies to both panels.

Owrutsky have also discussed the prospects for strong coupling in this pressure broadened regime.[3] Even higher Rabi splittings may be achievable using a cavity chamber designed to handle compressed gases, though resolving such splittings will require cavities with comparably large mode spacings more akin to the microcavities used for solution-phase polaritonics. Characterizing strong coupling under a range of gas pressures and temperatures will facilitate careful analysis of the disparate influences of homogeneous (pressure) and inhomogeneous (Doppler) broadening on polariton formation and behavior.[3, 47] This topic is of increasing interest as many groups are beginning to consider the role of disorder in cavity-altered processes.[43, 48-51] We hope that this first demonstration of room temperature gas-phase strong coupling lowers the barrier to entry for others interested in working in this domain.

### D. Exploring the impacts of cavity finesse on polariton formation

The cavity photon lifetime is a potentially important parameter in polariton chemistry,[52, 53] and one over which we have sensitive control via choice of cavity mirrors. Thus far, we have only reported on cavities with finesse $\mathcal{F} \sim 25$, using gold-coated mirrors with reflectivity $R \sim 88\%$. We initially chose these mirrors to yield cavity fringes comparable in linewidth to the molecular features of interest. We now examine the effects of changing cavity finesse on polariton formation. In Fig. 5a, we plot the transmission spectra of near-confocal cavities of length $L \sim$ 8.36 cm and finesses $\mathcal{F} \sim 10$, ~25 and ~43, with an intracavity number density $[CH_4] = 4.2 \times 10^{15}$ cm$^{-3}$ (5 sccm gas flow) at 115 K. We plot corresponding simulations of the cavity transmission spectra in Fig. 5b. Fig. S4 in the SM depicts the emergence of polariton peaks with increasing $CH_4$ flow rate for the $\mathcal{F} \sim 10$ and 43 cavities. We additionally show cavity dispersion plots for the $\mathcal{F} \sim 10$ cavity for two representative values of $[CH_4]$ in Fig. S5.

Cavities with finesses of $\mathcal{F} \sim 10$, 25, and 43 demonstrate photonic linewidths of $\Delta \nu = 166$, 70, and 54 MHz fwhm respectively, as observed for cavity fringes lying far from molecular absorption in Fig. 5a. These linewidths are consistent with the expected $\Delta \nu = \text{FSR}/\mathcal{F}$ values for a Fabry-Pérot cavity, where FSR is the free spectral range.[32] In each case, the polariton peaks inherit the homogeneous cavity linewidth, as others have observed.[47] The Rabi splitting additionally appears independent of the cavity finesse over the range of conditions we explore here. Although narrower cavity linewidths may concomitantly lower the Rabi splitting required to realize the strong coupling condition,[39] the effect of mismatched linewidths on $\Omega_R$ is negligible for a large number of emitters,[3] as is the case for our system. Here, we have exclusively used cavities constructed from gold mirrors with as low finesse as is practical to achieve VSC. In future efforts, it would be possible to use higher-reflectivity mirrors to engineer considerably higher-finesse systems as are commonly used in cavity-enhanced spectroscopy.[54] The ability to integrate cavities of widely varying finesse into our apparatus will allow us to match the photonic linewidth to a range of molecular lineshapes as well as to interrogate the influence of cavity losses on intracavity processes.[12, 55]

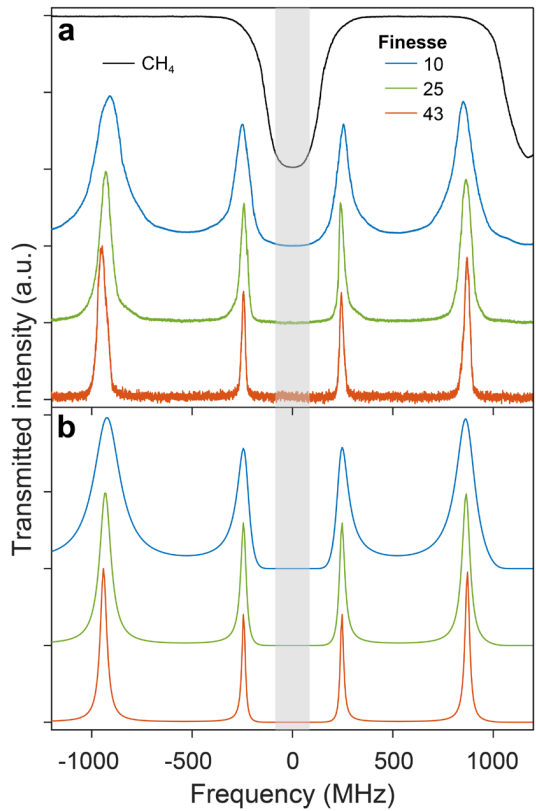

**FIG. 5. (a)** Experimental transmission spectra of near-confocal Fabry-Pérot cavities of length $L \sim 8.36$ cm and of varying finesse with an intracavity number density $[CH_4] = 4.2 \times 10^{15}$ cm$^{-3}$ (5 sccm gas flow) at ~115 K. Successive spectra are vertically offset for clarity, as is the experimental transmission spectrum of $CH_4$ acquired under the same experimental conditions (black). In all traces, a cavity mode is kept locked on resonance with the $CH_4$ $\nu_3$ $J = 3 \rightarrow 4$ $A_2(0)$ transition at 3057.687423 cm$^{-1}$. The frequency axis (300 MHz $\approx$ 0.01 cm$^{-1}$) is referenced with this transition corresponding to 0 MHz. **(b)** Simulated cavity transmission spectra obtained by fitting the corresponding experimental traces. The legend applies to both panels.

### E. Changing photonic mode density via optical cavity mode geometry

Thus far, we have limited our discussion to consider only plano-concave Fabry-Pérot cavities with near-confocal geometries where the mirror ROC is matched to the cavity length. To explore changing the photonic mode structure of our cavities, we have additionally considered strong coupling in the non-confocal case. While confocal cavities are straightforward to align and feature highly-degenerate transverse spatial modes, a non-confocal cavity geometry has the benefit of sparser cavity fringes, with effectively twice the mode spacing of a confocal cavity of the same length,[32, 56] which permits less congested cavity-coupling conditions.

In Fig. 6, we compare cavity transmission spectra for a non-confocal (Fig. 6a) and a near-confocal (Fig. 6b) cavity with one cavity fringe resonantly matched to the $CH_4$ $\nu_3$ $J = 3 \rightarrow 4$ $A_2(0)$ transition. The near-confocal data is reproduced from our previous work[26] where we used a 100 sccm flow of He buffer gas to cool the $CH_4$ to a temperature of 120 K; we use this same experimental configuration for the non-confocal cavity. Both cavities feature identical cavity length ($L \sim 8.36$ cm) and finesse ($\mathscr{F} \sim 25$), but feature mirror ROCs of $-21.69$ cm and $-8.36$ cm for the non-confocal and near-confocal cases, respectively. In both cases, the resonant cavity fringe is observed to broaden and then split as [$CH_4$] increases. The observed Rabi splitting for the near-confocal cavity is slightly larger than that of the non-confocal cavity by a factor of roughly 1.25 for a given [$CH_4$] (Fig. 6c). The non-confocal cavity also features polariton peaks which are consistently lower in intensity than those of the near-confocal cavity.

Since the non-confocal and near-confocal cavities are identical apart from their mirror ROC, we speculate that the differences in their transmission spectra must arise from their distinct photonic mode structures. While the planar microcavities used for solution-phase strong coupling feature a continuum of modes with finite in-plane momenta, our cavities support discrete spectra of transverse Gaussian spatial modes. Confocal cavities feature highly degenerate longitudinal and transverse modes, while non-confocal cavities have more congested spectra that depend on the ratio of cavity length to mirror ROC. The Rabi splitting in our near-confocal cavity (Fig. 6b) therefore arises from the coupling of the target molecular transition to many degenerate spatial cavity modes, while the splitting in the non-confocal case (Fig. 6a) derives principally from light-matter interaction with the TEM00 spatial mode to which the probe laser is mode-matched. Wickenbrock *et al.*[57] find that $\Omega_R$ is independent of the number of coupled degenerate cavity modes for a spatially uniform distribution of emitters, as each photonic mode independently couples to the intracavity ensemble with an identical $\Omega_R$. The slight difference in Rabi splitting that we observe for the non-confocal and near-confocal cases therefore remains a mystery. However, we might rationalize the difference in transmission intensity of the polariton features in the two cases in terms of inter-mode light scattering.[57] In brief, intracavity molecules can scatter light introduced in one spatial mode into other spatial modes. In the near-confocal case, scattered light can be recaptured by the degenerate spatial modes and ultimately transmitted through the cavity. In the non-confocal case, scattered light is more likely to be lost from the cavity transmission as there are no degenerate spatial modes to scatter into.

The spatial mode spectrum and density of states can be carefully engineered in open Gaussian cavities, unlike in planar microcavities. Such control is a crucial degree of freedom for carefully designed experiments in polariton chemistry and dynamics, as all orthogonal spatial modes can couple to an intracavity molecular ensemble, whether or not they are probed spectroscopically. In general, a more complete understanding the role of photonic mode density is important as we develop and test mechanistic theories for cavity-altered molecular phenomena.[5, 50]

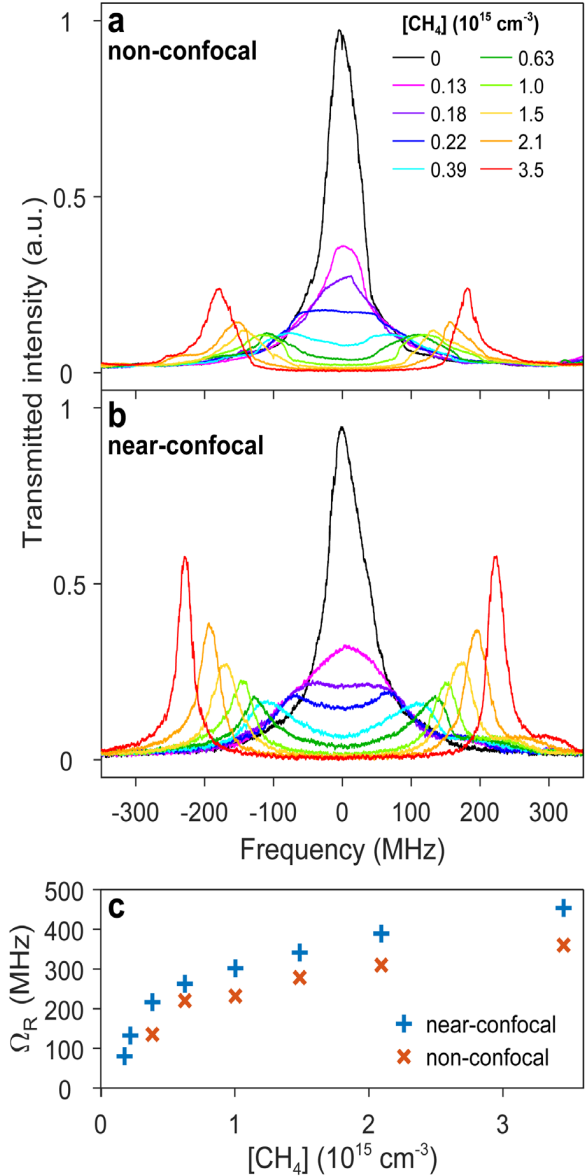

**FIG. 6. (a)** Experimental transmission spectra of a non-confocal Fabry-Pérot cavity with $\mathscr{F} \sim 25$, length $L \sim 8.36$ cm, and mirror ROC $=-21.69$ cm for various intracavity number densities [CH$_4$] at ~120 K. In all traces, a cavity mode is kept locked on resonance with the CH$_4$ $\nu_3$ $J = 3\rightarrow 4$ $A_2(0)$ transition at 3057.687423 cm$^{-1}$. The frequency axis (300 MHz ≈ 0.01 cm$^{-1}$) is referenced with this transition corresponding to 0 MHz. The traces are normalized against the intensity of a fringe far from resonance. **(b)** Corresponding experimental transmission spectra of a near-confocal Fabry-Pérot cavity of finesse $\mathscr{F} \sim 25$, length $L \sim 8.36$ cm, and mirror ROC $=-8.36$ cm for corresponding intracavity number densities at ~120 K. The legend applies to both panels (a) and (b). **(c)** The Rabi splitting, $\Omega_R$, extracted from the spectra in panels (a) and (b) and plotted as a function of [CH$_4$]. Panel (b) is reprinted with permission from Ref. 26, A. D. Wright, J. C. Nelson and M. L. Weichman, J. Am. Chem. Soc. **145**, 5982 (2023). Copyright 2023 American Chemical Society.

## IV. CONCLUSIONS AND FUTURE PROSPECTS

Over the past eight years, the molecular polariton community has examined different facets of the behavior of solution-phase molecules under VSC, exploring conditions spanning the weak to strong coupling regimes,[58] multimode coupling,[38] multi-transition coupling schemes,[44, 46] and ultimately vibrational polariton chemistry and dynamics.[2, 3, 5, 11, 13, 16, 59, 60] Here, we draw inspiration from this growing condensed-phase literature as we characterize the range of capabilities of our complementary platform for gas-phase vibrational polaritons.

Our gas-phase cavity occupies a very different experimental regime than the more typical planar microfluidic cavities. We achieve strong coupling despite our centimeter-scale cavity length being a factor of ~$10^4$ times longer than the target transition wavelengths, and our cavity mirrors encompassing a small solid angle from the perspective of intracavity molecules. We plan to move towards smaller, millimeter-scale cavities in future work, which will make it easier to optically pump and spectroscopically probe the entire cavity mode volume orthogonal to the cavity axis. Working with shorter cavities may also assist in future efforts to cavity-couple rovibrationally-resolved electronic transitions. Although the collective coupling strength between cavity and molecules does not change with cavity length for fixed intracavity number density,[5, 39, 58] the stronger per-molecule coupling strength in shorter cavities may also promote more prominent cavity-altered effects.[15] We have particular interest in targeting unimolecular reaction systems in which intramolecular vibrational redistribution plays a prominent role.[12, 13, 61, 62] Altogether, this platform is well-poised for future studies of intracavity dynamics in pristine benchmark systems.

## SUPPLEMENTARY MATERIAL

See the supplementary material for the following additional figures: Figure S1 depicts the laser power-independence of the cavity transmission spectrum. Figure S2 illustrates how we account for absorption from extracavity $CH_4$ in our simulated transmission spectra. Figure S3 compares cavity transmission spectra for cavities with differing lengths. Figures S4 and S5 depict additional experimental transmission spectra and dispersion plots for near-confocal cavities with varying finesse.

## ACKNOWLEDGEMENTS

Marissa L. Weichman acknowledges support from the ACS Petroleum Research Fund under project PRF-62543-DNI6, NSF CAREER award CHE-2238865, and startup funds provided by Princeton University. The authors acknowledge the use of Princeton's Imaging and Analysis Center (IAC), which is partially supported by the Princeton Center for Complex Materials (PCCM), a National Science Foundation (NSF) Materials Research Science and Engineering Center (MRSEC; DMR-2011750). This research made use of the Micro and Nano Fabrication Center (MNFC) at Princeton University.

## AUTHOR DECLARATIONS

**Conflict of Interest**

The authors have no conflicts to disclose.

**Author Contributions**

**Adam D. Wright:** Conceptualization; Data curation; Formal analysis; Investigation; Methodology; Visualization; Writing – original draft; Writing – review & editing

**Jane C. Nelson:** Formal analysis; Investigation; Methodology; Writing – review & editing

**Marissa L. Weichman:** Conceptualization; Funding acquisition; Methodology; Project administration; Resources; Supervision; Writing – review & editing

**Data availability**

The data that support the findings of this study are available from the corresponding author upon reasonable request.

# Supplementary Material:
# A versatile platform for gas-phase molecular polaritonics

*Adam D. Wright, Jane C. Nelson, Marissa L. Weichman\**

Department of Chemistry, Princeton University, Princeton, New Jersey 08544, United States

\* weichman@princeton.edu

## S1. Additional figures

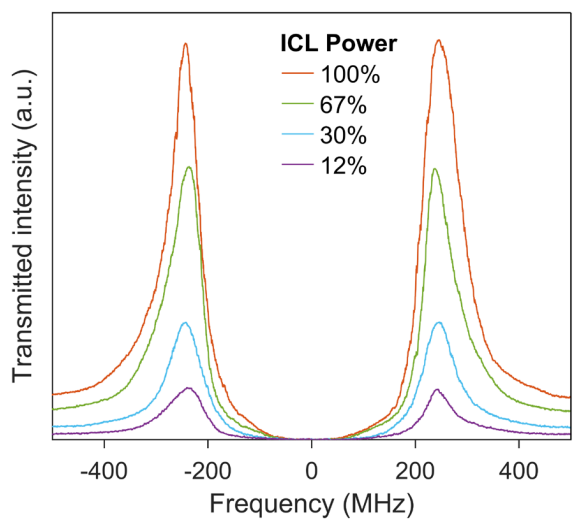

**FIG. S1.** Experimental transmission spectra of a near-confocal Fabry-Pérot cavity of finesse $\mathscr{F} \sim 10$ and length $L \sim 8.36$ cm, with $[CH_4] = 4.2 \times 10^{15}$ cm$^{-3}$ (5 sccm gas flow) at ~115 K. 100% laser power corresponds to 430 µW incident on the input window of the vacuum chamber. The cavity transmission spectrum remains consistent as the laser power is decreased. In all traces, a cavity mode is kept locked on resonance with the $CH_4$ $\nu_3$ $J = 3 \rightarrow 4$ $A_2(0)$ transition at 3057.687423 cm$^{-1}$. The frequency axis (300 MHz $\approx$ 0.01 cm$^{-1}$) is referenced with this transition corresponding to 0 MHz.

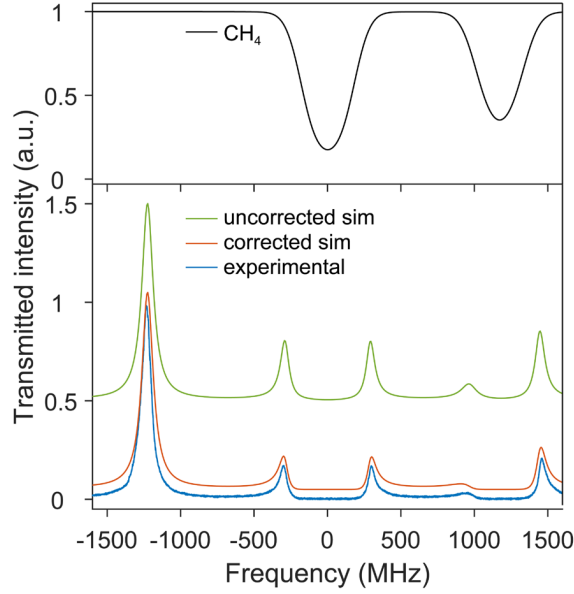

**FIG. S2.** Experimental transmission spectrum of a near-confocal Fabry-Pérot cavity of finesse $\mathcal{F} \sim 25$ and length $L \sim 6.27$ cm, acquired with $[CH_4] = 1.3 \times 10^{16}$ cm$^{-3}$ (200 sccm gas flow) at 295 K (blue). One cavity mode is kept locked on resonance with the $CH_4$ $v_3$ $J = 3 \rightarrow 4$ $A_2(0)$ transition at 3057.687423 cm$^{-1}$. The frequency axis is referenced with this transition corresponding to 0 MHz. The trace simulated using the uncorrected Eq. (1) of the main text (green) does not accurately reproduce the experimental intensities of the polariton features. The corrected simulation (orange), on the other hand, closely reproduces the experimental spectrum by accounting for the absorption of light by extracavity $CH_4$ along the beampath. The correction is applied by multiplying Eq. (1) by $e^{-\alpha(v)d}$, where $\alpha(v)$ is the frequency-dependent $CH_4$ absorption coefficient and $d$ is the pathlength over which extracavity $CH_4$ can absorb light. The single-pass transmission spectrum of $CH_4$ under these conditions is plotted for reference in black, calculated as $e^{-\alpha(v)L}$. Spectra are vertically offset for clarity.

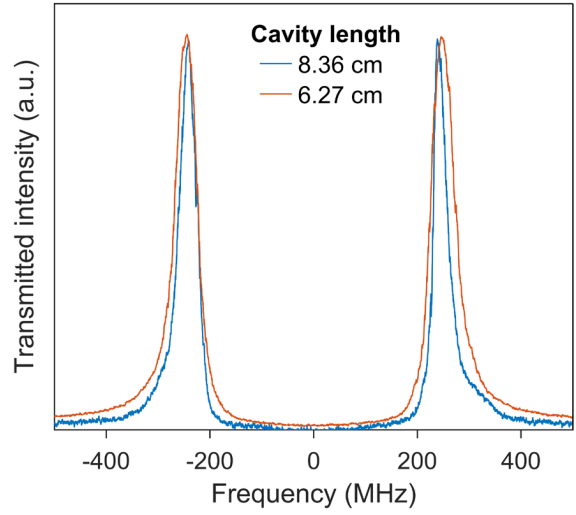

**FIG. S3.** Experimental transmission spectra of near-confocal Fabry-Pérot cavities of finesse $\mathscr{F} \sim 25$ and lengths $L \sim 8.36$ cm (blue) and $L \sim 6.27$ cm (orange) with $[CH_4] = 4.2 \times 10^{15}$ cm$^{-3}$ (5 sccm gas flow) at ~115 K. The Rabi splitting is not observed to change with cavity length, though reducing the cavity length increases the cavity linewidth, which the polariton peaks inherit. One cavity mode is kept locked on resonance with the $CH_4$ $v_3$ $J = 3 \rightarrow 4$ $A_2(0)$ transition at 3057.687423 cm$^{-1}$. The frequency axis is referenced with this transition corresponding to 0 MHz.

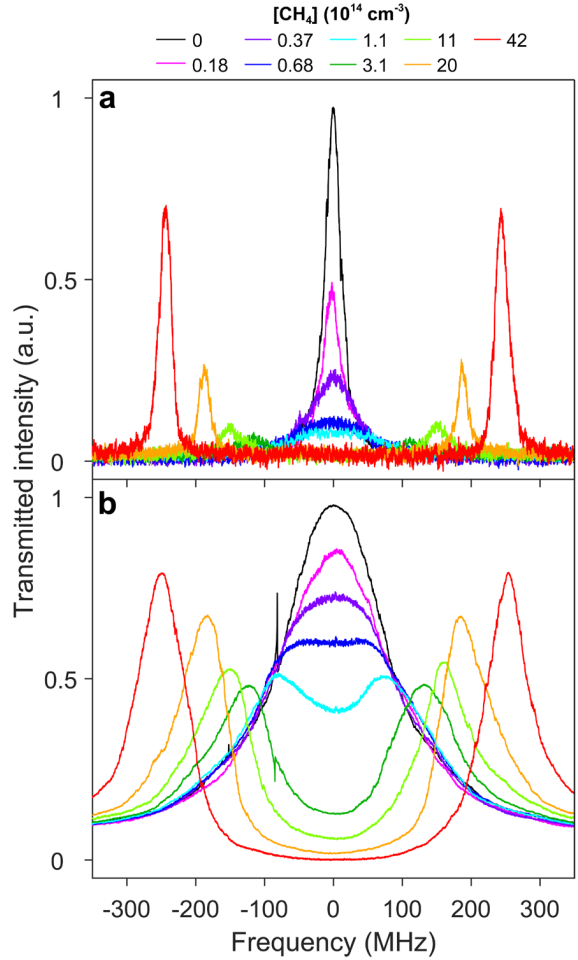

**FIG. S4.** Experimental transmission spectra of near-confocal Fabry-Pérot cavities of length $L \sim 8.36$ cm and finesses **(a)** $\mathscr{F} \sim 43$ and **(b)** $\mathscr{F} \sim 10$, for various intracavity number densities [CH$_4$] at ~115 K. The polariton linewidths reflect the cavity finesse in both cases, though the Rabi splitting appears independent of finesse over the range of number densities shown. In all traces, a cavity mode is kept locked on resonance with the CH$_4$ $\nu_3$ $J = 3 \rightarrow 4$ $A_2(0)$ transition at 3057.687423 cm$^{-1}$. The frequency axis (300 MHz $\approx$ 0.01 cm$^{-1}$) is referenced with this transition corresponding to 0 MHz. The legend applies to both panels.

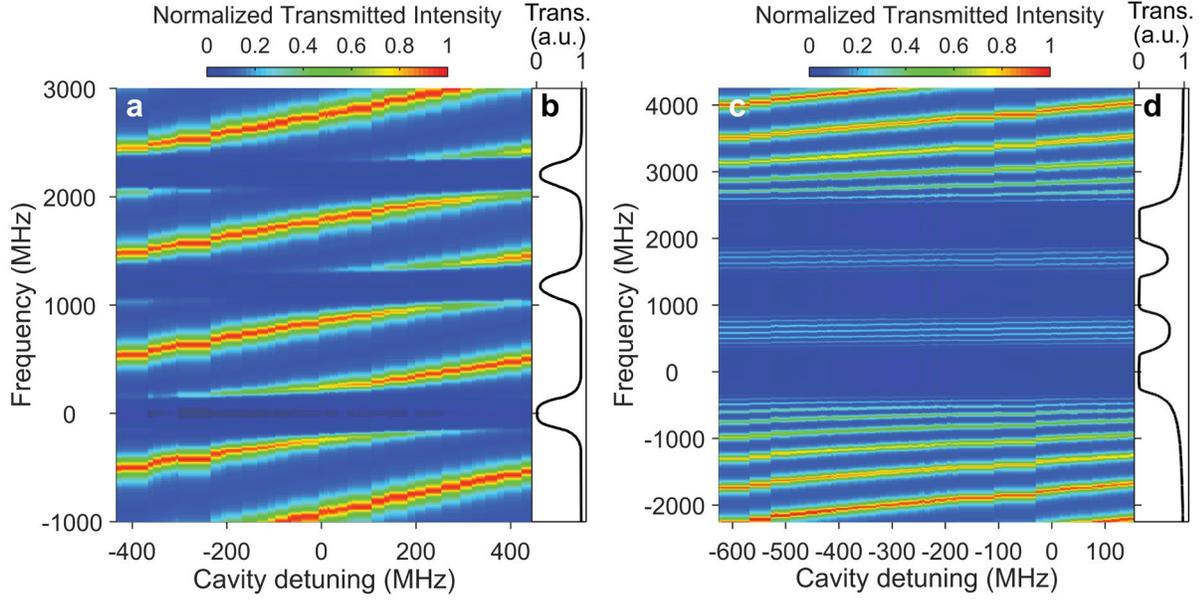

**FIG. S5.** Dispersion plots of the experimental transmission spectra of a near-confocal Fabry-Pérot cavity with finesse $\mathcal{F} \sim 10$ and length $L \sim 8.36$ cm, with intracavity number densities of of **(a)** $[CH_4] = 4.2 \times 10^{15}$ cm$^{-3}$ (5 sccm gas flow) and **(c)** $[CH_4] = 1.6 \times 10^{17}$ cm$^{-3}$ (1000 sccm gas flow) $CH_4$ at ~115 K. The cavity is systematically detuned from resonance by varying its length. The frequency axis (300 MHz $\approx 0.01$ cm$^{-1}$) is referenced with the $CH_4$ $\nu_3$ $J = 3 \rightarrow 4$ $A_2(0)$ transition at 3057.687423 cm$^{-1}$ corresponding to 0 MHz. **(b,d)** Experimental transmission spectra of $CH_4$ shown for reference under the same conditions as the corresponding dispersion plots. Note that cavity optical resonances which lie within the strongly absorbing regions are extinguished and so do not appear in the dispersion plots.